\documentclass{mem}
\usepackage{natbib}\usepackage{txfonts}\usepackage{balance}
\usepackage{graphicx}
\usepackage[breaklinks]{hyperref}
\idline{153}{498}
\begin{document}

\title{
Planetary systems in star clusters
}

   \subtitle{}

\author{M.B.N. \,Kouwenhoven\inst{1,2},
Qi\,Shu\inst{2,3},
Maxwell Xu\, Cai\inst{4,5,2},
\and Rainer\, Spurzem\inst{5,2,6}}

\institute{
Department of Mathematical Sciences, Xi'an Jiaotong-Liverpool University, 111 Ren'ai Road, Dushu Lake Higher Education Town, Suzhou Industrial Park, Suzhou, 215123, P.R. China \email{t.kouwenhoven@xjtlu.edu.cn}
\and
Kavli Institute for Astronomy and Astrophysics, Peking University, Yiheyuan Lu 5, Haidian Qu, Beijing 100871, P.R. China 
\and
Department of Astronomy, School of Physics, Peking University, Yiheyuan Lu 5, Haidian Qu, Beijing 100871, P.R. China 
\and
Leiden Observatory, Leiden University, PO Box 9513, 2300 RA, Leiden, The Netherlands
\and
National Astronomical Observatories of China and Key Lab for Computational Astrophysics, Chinese Academy of Sciences, 20A Datun Road, Chaoyang District, Beijing 100012, P.R. China
\and
Astronomisches Rechen-Institut, Zentrum f\"ur Astronomie, University of Heidelberg, M\"onchhofstrasse 12-14, 69120 Heidelberg, Germany
}

\authorrunning{Kouwenhoven, Shu, Cai \& Spurzem}

\titlerunning{Planetary systems in star clusters}

\abstract{
Thousands of confirmed and candidate exoplanets have been identified in recent years. Consequently, theoretical research on the formation and dynamical evolution of planetary systems has seen a boost, and the processes of planet-planet scattering, secular evolution, and interaction between planets and gas/debris disks have been well-studied. Almost all of this work has focused on the formation and evolution of isolated planetary systems, and neglect the effect of external influences, such as the gravitational interaction with neighbouring stars. Most stars, however, form in clustered environments that either quickly disperse, or evolve into open clusters. Under these conditions, young planetary systems experience frequent close encounters with other stars, at least during the first $10^6-10^7$ years, which affects planets orbiting at any period range, as well as their debris structures.
}
\maketitle{}


\section{Introduction}
\vspace{-0.2cm}

Observations have shown that young stars are almost exclusively found in groups, suggesting that stars form in clustered environments. Following the star formation process and gas expulsion phase, these stellar groupings may either virialise within several crossing times \citep[e.g.,][]{allison2009} and evolve in to star clusters that may survive for hundreds of millions of years, or disperse within several tens of millions of years \citep[e.g.,][]{degrijs2007}.
These observations, together with isotopic ratios in meteorites that indicate that a supernova explosion occurred within $\sim 1$~pc from the circumsolar disk, strongly suggests that also our Solar system formed in a clustered environment \citep[e.g.,][and references therein]{pfalzner2013}. Our Solar system, as well the thousands of known exoplanet systems, were likely born in environments where close stellar encounters were frequent, and their dynamical architectures may have been affected by such encounters. The major uncertainty, of course, is how important the effect of those encounters was, and for how long these stars remained part of this environment. As exoplanets appear to be substantially more common in the Solar neighbourhood than in star clusters, this indeed suggests that close stellar encounters have a destructive effect on planetary systems, unless the planet formation process is less efficient in environments that result in long-lived star clusters. In this article we provide a brief review on our strategy to answer these issues.


\vspace{-0.2cm}

\section{Single-planet systems in  clusters}

Observations, as well as the core-accretion scenario and the disk-fragmentation scenario for the formation of planetary-mass objects predict the formation of multiple planets, rather than single-planet systems, in circumstellar disks \citep[e.g.,][]{zhou2007, stamatellos2011, xie2014}. This multiplicity introduces planet-planet interactions and contribute to the fragility of these systems. Nevertheless, it is valuable to study single-planet systems, as these provide upper limits for the stability of such systems.
Single-planet systems in star clusters are substantially easier to model than multi-planet systems (see \S\ref{section:multiplanet}), and therefore much of the earlier work has focused on this approach. 
A comprehensive study of singe-planet systems in clusters, involving both $N$-body simulations and Monte Carlo simulations, was carried out by \cite{spurzem2009}. In their study of single-planet systems in large ($N=300\,000$) clusters, they find that, as expected, close (nearly parabolic, adiabatic) encounters are responsible for many of the disruptions. They also stress, however, that the cumulative effect of numerous weak (hyperbolic adiabatic) encounters can lead to substantial changes in the orbital elements of planets, which may result orbit crossing and subsequent decay of multi-planet systems.
\cite{parker2012} and \cite{zheng2015} modelled single-planet systems in star clusters, with the aim of obtaining relations for survival rates in different environments and for different orbital periods. \cite{zheng2015} obtain function fits for the survival and decay rates of single-planet systems in different  cluster environments. As multiplicity decreases survival rates of planetary systems, these  provide upper limits for planetary survival rates.


\vspace{-0.2cm}

\section{Free-floating planets in  clusters}

Following the escape from their host stars, free-floating planets (FFPs) become part of the star clusters dynamics. FFPs with velocities above the local escape velocity may escape immediately, while the others remain part of the star clusters, where they have a small chance of being re-captured by another star \citep[e.g.,][]{kouwenhoven2010, perets2012}, but the majority ultimately escapes from the star cluster through ejection or evaporation \citep[e.g.,][]{liu2013}. \cite{wang2015a} carried out an extensive study on the behaviour of free-floating planets in star clusters, and find that FFPs (with ejection velocities below the local escape velocity) typically remain part of the cluster for long times, and experience several (sometimes up to hundreds of) close ($<1000$~AU) encounters with neighbouring stars. Strong scattering ejects FFPs immediately, while weak scattering results in the gradual migration of FFPs to the outer regions of the cluster, where they are gently pruned off by the Galactic tidal field. At any time, a FFP has statistically a 40\% larger chance of being ejected than a typical star. FFPs are thus mostly ejected at early times, although several remain bound to the cluster until its dissolution. The linearly decreasing trend of the FFP-to-star ratio among cluster members found by \cite{wang2015a} indicates that surveys for free-floating planets in star clusters are will likely be most successful in young environments.


\vspace{-0.2cm}

\section{Multi-planet systems in clusters} \label{section:multiplanet}

Modelling multi-planet systems in star clusters is complex due to the enormous ranges in masses, positions, and velocities. Round-off errors in the calculation of the dynamical properties of stars are generally disastrous for modelling the evolution of planetary systems \citep[except for certain special cases which can be modelled as perturbed two-body or three-body systems; e.g., ][]{shara2016}. The study of multi-planet systems in star clusters has been hampered by these computational difficulties for long. \cite{malmberg2011} and \cite{hao2013} approximated the situation by modelling the close encounters with stars that occur in a star cluster, and carrying out separate simulations using {\tt MERCURY6} \citep{chambers1999}, instead of carrying out full star cluster simulations with multi-planet systems. \cite{malmberg2011} noted that a close encounter can result in the ejection of planets long (up to tens of millions of years) after the encounter occurred. \cite{hao2013} found that even close-in planets, which are themselves not directly affected by close encounters, may experience a collision or ejection due to planet-planet scattering. \cite{hao2013} also stressed the importance of the planetary mass spectrum. When perturbing the outermost planets of our own Solar system, for example, Jupiter's large inertia almost guarantees dynamical protection of the inner Solar system. With the newly implemented HDF5 storage scheme in {\tt NBODY6++} \citep{cai2015} and the development of the AMUSE framework \citep[e.g.,][]{portegies2013, pelupessy2013}, it recently became possible to combine the planetary dynamics code {\tt MERCURY6} with the star cluster code {\tt NBODY6++} \citep{spurzem1999} in {\tt AMUSE} \citep{cai2016}. In an upcoming publication (Cai et al., in prep.) we will outline the diverse dynamical outcomes of perturbed multi-planet systems in star clusters, confirming the importance of the interplay between the effect of encounters on  outer planets and the subsequent planet-planet scattering and secular interactions, and also confirming \cite{hao2013}'s findings that even the closest-in planets are (indirectly) affected by stellar encounters.


\vspace{-0.2cm}

\section{Debris structures in star clusters}

Our Solar system contains billions of comets, Kuiper Belt objects (KBOs) and asteroids. There is no reason to expect that other stars lack similar debris systems. In crowded stellar environments the analogs of the Kuiper Belt and Oort Cloud may be substantially altered by close encounters \citep[e.g.,][]{brasser2015}. As comets affect the prospects for life at planets in the habitable zone, both destructively (impacts) and constructively (water delivery), modelling the evolution of circumstellar debris in star cluster environments is particularly interesting. 
 These processes are particularly important when the stellar density is high, such as in open or globular clusters. 
Over recent years, $N$-body simulation software for modelling star clusters has been significantly improved, allowing us to model globular clusters containing more than a million stars over a Hubble time using the {\tt NBODY6++GPU} code \citep{wang2015b, wang2016}. This code is highly optimised for star cluster simulations and very fast. This is exactly the reason why it is not possible to use {\tt NBODY6++GPU} for star clusters with massless particles such as comets, as many of these optimisations, such as KS-regularisation, the neighbour scheme, and individual time steps, are not suitable for large numbers of bodies with mass ratios approaching zero. For this reason, a new release of the code, {\tt NBODY6++GPU-MASSLESS}, is now under development (Shu et al, in prep). This code will be able to run simulations of star clusters containing hundreds of thousands of stars and massless particles, allowing us to study the evolution of Oort Clouds, Kuiper Belts, asteroid belts, and planetesimal disks in crowded stellar environments, as well as free-floating comets, asteroids, and planetesimals.


\vspace{-0.2cm}

\section{Summary}

Most planetary systems spend the first $10^6-10^7$ years of their existence in crowded stellar environments, while a small fraction reside in open cluster or globular clusters for up to billions of years. Gravitational interactions with neighbouring stars shape these planetary systems, leading to orbital reconfigurations, planet-planet scattering events, ejections, and physical collisions. Debris structures, consisting of comets, asteroids, KBOs, and planetesimals, are also shaped by these stellar encounters. We have carried out $N$-body simulations of planetary systems and free-floating planets in star clusters, and present our first results above.
Despite the enormous advances that were made using $N$-body codes, many physical processes are still difficult to model, such as the effect of gas dynamics and stellar feedback that is often seen in star-forming regions \citep[e.g.,][]{bik2010}, the inclusion of the processes of core-accretion and disk-fragmentation that are responsible for planet formation \citep[e.g.,][]{li2015, li2016}, combined with very large fractions of primordial stellar binaries \citep[e.g.,][]{kouwenhoven2005, kouwenhoven2007, kouwenhoven2009}. With the development of new software and hardware, realistic simulations of planetary systems, from birth to old age,   will become possible in the near future.


\vspace{-0.1cm}
 
\begin{acknowledgements}
 M.B.N.K. was supported by the Peter and Patricia Gruber Foundation, by the Peking University One Hundred Talent Fund (985), and by the National Natural Science Foundation of China (grants 11010237, 11050110414, 11173004, 11573004). This publication was made possible through the support of a grant from the John Templeton Foundation and the National Astronomical Observatories of Chinese Academy of Sciences. We acknowledge support by Chinese Academy of Sciences through the Silk Road
Project at NAOC, through the Chinese Academy of Sciences Visiting Professorship for Senior International Scientists, Grant Number 2009S1-5 (R.S.), and through the ÒQianrenÓ special foreign experts program of China.
The special GPU accelerated supercomputer {\tt laohu} at the Center of Information and
Computing at National Astronomical Observatories, Chinese Academy of Sciences, funded by Ministry of Finance of PeopleÕs Republic of China under the grant ZDY Z2008-2, has been used for some simulations. 
\end{acknowledgements}

\vspace{-0.5cm}

\bibliographystyle{aa}

\begin{thebibliography}{}



\bibitem[Allison et al.(2009)]{allison2009} Allison, R.~J., Goodwin, S.~P., Parker, R.~J., et al.\ 2009, \apjl, 700, L99 

\bibitem[Bik et al.(2010)]{bik2010} Bik, A., Puga, E., Waters, L.~B.~F.~M., et al.\ 2010, \apj, 713, 883 

\bibitem[Brasser \& Schwamb(2015)]{brasser2015} Brasser, R., \& Schwamb, M.~E.\ 2015, \mnras, 446, 3788 

\bibitem[Cai et al.(2015)]{cai2015} Cai, M.~X., Meiron, Y., Kouwenhoven, M.~B.~N., Assmann, P., \& Spurzem, R.\ 2015, \apjs, 219, 31 

\bibitem[Cai et al.(2016)]{cai2016} Cai, M.~X., Spurzem, R., \& Kouwenhoven, M.~B.~N.\ 2016, IAU Symposium, 312, 235 

\bibitem[Chambers(1999)]{chambers1999} Chambers, J.~E.\ 1999, \mnras, 304, 793 

\bibitem[de Grijs \& Parmentier(2007)]{degrijs2007} de Grijs, R., \& Parmentier, G.\ 2007, ChJAA, 7, 155 

\bibitem[Hao et al.(2013)]{hao2013} Hao, W., Kouwenhoven, M.~B.~N., \& Spurzem, R.\ 2013, \mnras, 433, 867 

\bibitem[Kouwenhoven et al.(2005)]{kouwenhoven2005} Kouwenhoven, M.~B.~N., Brown, A.~G.~A., Zinnecker, H., Kaper, L., \& Portegies Zwart, S.~F.\ 2005, \aap, 430, 137 


\bibitem[Kouwenhoven et al.(2007)]{kouwenhoven2007} Kouwenhoven, M.~B.~N., Brown, A.~G.~A., Portegies Zwart, S.~F., \& Kaper, L.\ 2007, \aap, 474, 77 

\bibitem[Kouwenhoven et al.(2009)]{kouwenhoven2009} Kouwenhoven, M.~B.~N., Brown, A.~G.~A., Goodwin, S.~P., Portegies Zwart, S.~F., \& Kaper, L.\ 2009, \aap, 493, 979 

\bibitem[Kouwenhoven et al.(2010)]{kouwenhoven2010} Kouwenhoven, M.~B.~N., Goodwin, S.~P., Parker, R.~J., et al.\ 2010, \mnras, 404, 1835 

\bibitem[Li et al.(2015)]{li2015} Li, Y., Kouwenhoven, M.~B.~N., Stamatellos, D., \& Goodwin, S.~P.\ 2015, \apj, 805, 116 

\bibitem[Li et al.(2016)]{li2016} Li, Y., Kouwenhoven, M.~B.~N., Stamatellos, D., \& Goodwin, S.~P.\ 2016, arXiv:1609.00120 

\bibitem[Liu et al.(2013)]{liu2013} Liu, H.-G., Zhang, H., \& Zhou, J.-L.\ 2013, \apj, 772, 142 

\bibitem[Malmberg et al.(2011)]{malmberg2011} Malmberg, D., Davies, M.~B., \& Heggie, D.~C.\ 2011, \mnras, 411, 859 

\bibitem[Parker \& Quanz(2012)]{parker2012} Parker, R.~J., \& Quanz, S.~P.\ 2012, \mnras, 419, 2448 


\bibitem[Pelupessy et al.(2013)]{pelupessy2013} Pelupessy, F.~I., van Elteren, A., de Vries, N., et al.\ 2013, A\&A, 557, A84 

\bibitem[Perets \& Kouwenhoven(2012)]{perets2012} Perets, H.~B., \& Kouwenhoven, M.~B.~N.\ 2012, \apj, 750, 83 

\bibitem[Pfalzner(2013)]{pfalzner2013} Pfalzner, S.\ 2013, \aap, 549, A82 

\bibitem[Portegies Zwart et al.(2013)]{portegies2013} Portegies Zwart, S. et al. 2013, Computer Physics Communications, 183, 456

\bibitem[Shara et al.(2016)]{shara2016} Shara, M.~M., Hurley, J.~R., \& Mardling, R.~A.\ 2016, \apj, 816, 59 

\bibitem[Spurzem et al.(2009)]{spurzem2009} Spurzem, R., Giersz, M., Heggie, D.~C., \& Lin, D.~N.~C.\ 2009, ApJ, 697, 458

\bibitem[Spurzem(1999)]{spurzem1999} Spurzem, R.\ 1999, Journal of Computational and Applied Mathematics, 109, 407 


\bibitem[Stamatellos et al.(2011)]{stamatellos2011} Stamatellos, D., Whitworth, A.~P., \& Hubber, D.~A.\ 2011, \apj, 730, 32 

\bibitem[Wang et al.(2015a)]{wang2015a} Wang, L., Kouwenhoven, M.~B.~N., Zheng, X., Church, R.~P., \& Davies, M.~B.\ 2015a, \mnras, 449, 3543 

\bibitem[Wang et al.(2015b)]{wang2015b} Wang, L., Spurzem, R., Aarseth, S., et al.\ 2015b, \mnras, 450, 4070 

\bibitem[Wang et al.(2016)]{wang2016} Wang, L., Spurzem, R., Aarseth, S., et al.\ 2016, \mnras, 458, 1450 

\bibitem[Xie et al.(2014)]{xie2014} Xie, J.-W., Wu, Y., \& Lithwick, Y.\ 2014, \apj, 789, 165 

\bibitem[Zheng et al.(2015)]{zheng2015} Zheng, X., Kouwenhoven, M.~B.~N., \& Wang, L.\ 2015, \mnras, 453, 2759 

\bibitem[Zhou et al.(2007)]{zhou2007} Zhou, J.-L., Lin, D.~N.~C., \& Sun, Y.-S.\ 2007, \apj, 666, 423 


\end{thebibliography}

\end{document}